\documentclass[conference,10pt]{IEEEtran}
\usepackage{epsfig,epsf,rotating,setspace,latexsym,amsmath,amssymb,amsfonts,bm,theorem,subfigure,epstopdf}
\usepackage{cite,authblk}
\usepackage{bbm}
\usepackage{algorithm}
\usepackage[noend]{algpseudocode}
\usepackage{color}
\usepackage{mathtools}
\usepackage{soul}
\DeclareMathOperator*{\argmax}{arg\,max}

\algrenewcommand\algorithmicforall{\textbf{foreach}}
\algrenewcommand\algorithmicindent{.8em}

\newtheorem{definition}{Definition}

\newtheorem{remark}{Remark}

\IEEEoverridecommandlockouts
\allowdisplaybreaks

\begin{document}

\title{How to Make Money From Fresh Data: Subscription Strategies in Age-Based Systems \thanks{Research of UIUC authors was supported in part by the ARO MURI Grant AG285 and the AFOSR Grant FA9550-19-1-0353.}}

\author[1]{Priyanka Kaswan}
\author[2]{Melih Bastopcu}
\author[1]{Sennur Ulukus}
\author[2]{S. Rasoul Etesami}
\author[2]{Tamer Ba{\c s}ar}

\affil[1]{\normalsize Department of Electrical and Computer Engineering, University of Maryland, College Park, MD 20742}
\affil[2]{\normalsize Coordinated Science Laboratory, University of Illinois Urbana Champaign, Urbana, IL-61801}

\maketitle

\begin{abstract}
We consider a communication system consisting of a server that tracks and publishes updates about a time-varying data source or event, and a gossip network of users interested in closely tracking the event. The timeliness of the information is measured through the version age of information. The users wish to have their expected version ages remain below a threshold, and have the option to either rely on gossip from their neighbors or subscribe to the server directly to follow updates about the event if the former option does not meet the timeliness requirements. The server wishes to maximize its profit by increasing the number of subscribers and reducing costs associated with the frequent sampling of the event. We model the problem setup as a Stackelberg game between the server and the users, where the server commits to a frequency of sampling the event, and the users make decisions on whether to subscribe or not. As an initial work, we focus on directed networks with unidirectional flow of information and obtain the optimal equilibrium strategies for all the players. We provide simulation results to confirm the theoretical findings and provide additional insights.
\end{abstract}

\section{Introduction} \label{sec:intro}
As we transition into the 5G/6G era, we are experiencing a surge in real-time applications where the value of data is tied to its freshness. Examples of such applications could be navigation systems in vehicular networks, environmental monitoring systems, financial markets, or simply news channels that cater to people's desire to stay informed about the latest events and developments in their local area. Maintaining fresh data, however, can be rather expensive, as it involves frequent sampling of the underlying time-varying processes or events, in addition to processing and transmission-related operational costs. As a result, freshness becomes a valuable commodity in the commercial realm, where IoT service providers or news agencies can monetize these services by pricing the delivery of fresh information \cite{zhang_pricingfreshdata}. 

In this respect, we examine the profit maximization strategy at such a service provider (or simply a server), which tracks and publishes updates about a continually updating process or event $E$. A set of users $\mathcal{N}$ wishes to stay closely informed about the latest updates regarding the event and can explore two avenues: they can either directly subscribe to the server for timely updates, or they can rely on intermittent updates, i.e., gossip from their neighboring users, as long as this approach ensures that they remain reasonably informed about the event. In the latter case, the timeliness of a user would depend on the underlying network topology, and the multiplicity and proximity of subscribing nodes.

\begin{figure}[t]
\centerline{\includegraphics[width=0.50\linewidth]{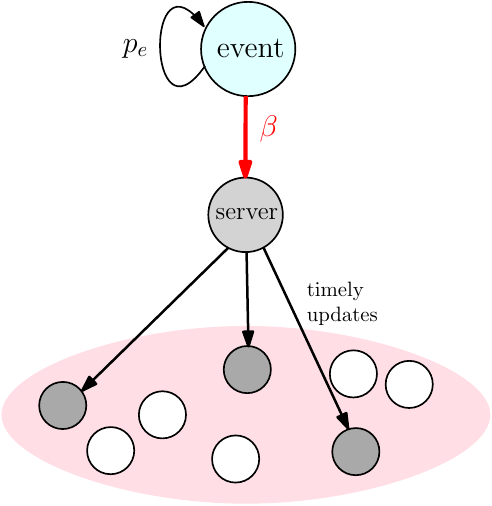}}\vspace*{-0.4cm}
\caption{System model consisting of a server and user nodes. Here, the event process is updated with probability $\!p_e$ and the server gets updates with rate $\!\beta\!$.}
\label{fig:system_model}
\vspace*{-0.5cm}
\end{figure}

Gossiping is a decentralized information dissemination mechanism, that has become popular due to its simplicity, scalability, and low maintenance costs, in which each user communicates with its neighbors probabilistically during each timeslot\cite{aoi_gossip_litsurvy, Yates21gossip, kaswan_jamming_jrl, arunabh_grid , arunabh_genring, baturalp21comm_struc, mitra_srcrate_learning24}. Prior works on gossip-based publish-subscribe systems \cite{ferretti_publishsubsgossp, persu_netwrk} have focused on a fixed message dissemination, as opposed to dynamic information dissemination as in this work. For instance, \cite{ferretti_publishsubsgossp} analyzes the average number of interested users to which a fixed message percolates in random networks built using Poisson degree distribution. In \cite{persu_netwrk}, the server carefully manipulates a message to persuade users into taking a desired action without disincentivizing users from subscribing, particularly when the gossiping users might default to an undesirable action in the absence of such a message. In these works, with probability $p$, each user forwards the message to a neighbor, which is modeled using the Erd\H{o}s-R\'enyi random graph approach where a link of the gossip network disappears with probability $1-p$. This prior line of work in \cite{ferretti_publishsubsgossp, persu_netwrk} has important differences from real-time network models, where each user forwards its most recently received version to a neighbor with probability $p$ in every timeslot, and thus, a particular version could potentially have multiple forwarding attempts on the same link. Further, under fixed message dissemination, a high gossip probability can ensure that a message reaches all users from only one subscribing user. However, a single subscriber might not be good enough for real-time systems since a non-subscriber situated far away from this subscriber may constantly receive outdated information due to multiple hops in information dissemination path, forcing it to opt for the first option of directly subscribing to the server. Hence, the portion of subscribers will be higher in real-time applications.

The set of questions we wish to investigate in this work is  the following: Given a freshness-based gossip network, which users will choose to subscribe to the server, and how often should the server sample the event process to sustain a desired fraction of subscribing users from the population while keeping the cost of sampling the process at a minimum in order to maximize its overall profit? The asymmetry in these two objectives leads to a non-cooperative game between the server and the users. Specifically, we model this as a Stackelberg game, where the server acts as the leader by committing to a sampling strategy for the event process. The users, on the other hand, act as followers who decide whether to subscribe or not to the server based on the server's chosen sampling strategy. We formalize the system model and study equilibrium subscription strategies for a range of gossip network topologies. We start with simpler network topologies, so that the results established in these initial topologies would be helpful in the analysis of more complex topologies later. Finally, we provide simulation results to confirm the theoretical findings of this paper and provide additional insights.

\section{System Model and Age Analysis}
We consider a communication system consisting of a server node $R$ that samples a real-time data source or event $E$ that is continuously updated and a set of user nodes $\mathcal{N} \subseteq \mathbb{Z}$ arranged into a gossip network; see Fig.~\ref{fig:system_model}. We represent the gossip network of user nodes through the directed graph $ \overrightarrow{G} =(\mathcal{N},\overrightarrow{E})$ with edge set $\overrightarrow{E}$, such that node $i$ gossips with node $j$ if there is an edge $ e_{ij}\in\overrightarrow{E}$. Time is divided into discrete time slots, such that the event gets updated in each timeslot with probability $p_e$, with the time-interval between two successive event updates distributed as a geometric distribution with parameter $p_e$. We employ the version age of information metric to characterize timeliness at nodes in this work as we associate a version number to every information about the event. Denoting the version number corresponding to the current state of the event by $V_E(t)$, each event update increments $V_E(t)$ by $1$. Similarly, we denote the version number of the information present at node $i$ by $V_i(t)$ and define the instantaneous version age at node $i$ at time $t$ as $X_i(t)=V_E(t)-V_i(t)$, capturing the notion of how much the user lags behind the state-of-the-art version of the event.

The server node gets an update about the event process with probability $\beta$ during each timeslot $t$, resetting its version number $V_R(t+1)= V_E(t)$ at the beginning of the following timeslot post update reception. Let $U_{i,j}(t)$ indicate whether an update packet is transmitted from node $i$ to node $j$ at time $t$: $U_{i,j}(t)=1$ implies that an update was sent such that the transmission consumed the whole timeslot $t$ and was available to the receiving node $j$ at the beginning of timeslot $t+1$, and $U_{i,j}(t)=0$ implies that no packet was sent. Analogously, we use $U_{E,R}(t)$ and $U_{E}(t)$ to indicate whether an update was received at the server and a new event update occurred, respectively. When $U_{E}(t)=1$, that is when a new event update occurs during timeslot $t$, the instantaneous version of all other nodes increases by $1$ in the next timeslot $t+1$ due to unit increment in $V_E(t+1)$ at the event. 

\subsection{Average Age Analysis at the Server}
When an update packet arrives at the server at time $t$, the instantaneous version age at the server $X_R(t)=V_E(t)-V_R(t)$ drops to $0$ in the next timeslot ($t+1$) in the absence of any event update. However, if a new event update simultaneously occurs at timeslot $t$, we assume that this update takes the whole timeslot to process and is not ready to be forwarded to the server within the same timeslot, in which case the instantaneous version age at the server increases by $\!1\!$ in the next timeslot. In the absence of any event updates or packet arrivals at the server, the instantaneous version age at the server remains unchanged. Hence, the four combinations resulting from transitions $U_{E}(t)$ and $U_{E,R}(t)$ cause $X_{R}(t)$ to evolve as 
\begin{align}\label{eqn:instantage_server}
    X_R(t+1) = & 0 \times (1-U_{E}(t))U_{E,R}(t) 
    + 1 \times U_{E}(t)U_{E,R}(t) \nonumber\\
    & + X_R(t) \times (1-U_{E}(t))(1-U_{E,R}(t))\nonumber\\
    & + (X_R(t)+1)\times U_{E}(t)(1-U_{E,R}(t)).
\end{align}
It is clear that (\ref{eqn:instantage_server}) together with some initial distribution defines a discrete time Markov chain with state space $\mathbb{Z}_{+}$. Showing that the Markov chain consists of single irreducible countably infinite states that is non-periodic and also positive recurrent is straightforward given the Bernoulli strategies on all update links, and we skip these details since they are non-instructive.

Define $x_R\!\!=\!\! \lim_{t \to \infty\!} \mathbb{E}[X_R(t)]$. Note that $U_{E}(t)$ and $U_{E,R}(t)$ are independent random variables, with expectations $p_e$ and $\beta$. Taking the long-term expectation on both sides of (\ref{eqn:instantage_server}) gives
\begin{align}\label{eqn:longtermexp_age_server}
    x_R=& 0 \times (1-p_e)\beta + 1\times p_e\beta + x_R \times (1-p_E)(1-\beta) \nonumber \\
    & + (x_R+1)p_e(1-\beta),
\end{align}
which simplifies to $x_R= p_e \beta^{-1}$.

When a user subscribes to the server, the server transmits the latest version it has at the beginning of each timeslot $t$ to the subscribing user (referred to as the subscriber), such that the subscriber receives the complete packet by the start of the next timeslot $t+1$. Therefore, the instantaneous version age at a subscriber, denoted by $X_S(t)$ (with $x_S=\lim_{t \to \infty} \mathbb{E}[X_S(t)]$),  is influenced by the version age at the server and any event updates in the preceding timeslot. Mathematically, this can be expressed as follows
\begin{align}\label{eqn:instantage_subscriber}
    X_S(t+1)= & X_R(t) \!\times\! (1\!-\!U_E(t) )  \!+\!(X_R(t)\!+\!1) \!\times\! U_E(t).
\end{align}
Taking long-term expectation and using $x_R=p_e \beta^{-1}$, we get
\begin{align}\label{eqn:longtermexp_age_subscriber}
    x_S= x_R+p_e = p_e\left(\beta^{-1}+1\right).
\end{align}

Now, we state the subscribing condition for the users. User $i$ prefers a bounded expected age $x_i=\lim_{t \to \infty} \mathbb{E}[X_i(t)]$, i.e.,
\begin{align}\label{eqn:ACconstraint}
    x_i<Lx_S,
\end{align}
for some $L\geq 1$, which we will call the \emph{age compatibility (AC) constraint}. If the condition in (\ref{eqn:ACconstraint}) is satisfied without subscription, user $i$ will not subscribe to the server, i.e., its subscription action will be $a_i =0$; otherwise, user $i$ will subscribe to the server, i.e., its subscription action will be $a_i =1$. When a user is a non-subscriber, it can still maintain timely information by requesting packets from its neighbors in the network, which implies that the network topology determines the user's action to subscribe or not subscribe. In a disconnected network, where each user is isolated from others, every user will decide to subscribe because non-subscription would result in an infinite expected age at the user, as they have no alternative way of getting updates other than from the server. For a general network structure, the following definition will be useful in identifying equilibrium strategies.

\begin{definition}\label{Defn:stable}
For a given network structure and the server's update rate $\beta$, we say that a user is AC-stable if the user
\begin{itemize}
    \item is a non-subscriber and its expected age satisfies (\ref{eqn:ACconstraint}); or
    \item  is a subscriber and the alternate decision to unsubscribe while keeping other users' decisions unchanged would cause its expected age to violate (\ref{eqn:ACconstraint}).
\end{itemize}  
\end{definition}

\begin{remark}
    Even though there is no cost associated with the subscription, due to the definition of AC-stable equilibrium, the users prefer not to subscribe if they can satisfy their age requirement through their neighbors.
\end{remark}

Thus, when the users are \textit{AC-stable,} no user wants to deviate from their subscription decisions. Next, we look at certain network topologies to understand server-preferred subscription equilibrium strategies defined as, for a given strategy of the server, among all AC-stable equilibrium strategies, users are assumed to follow the one that will work best for the server.

\begin{figure}[t]
\centerline{\includegraphics[width=0.80\linewidth]{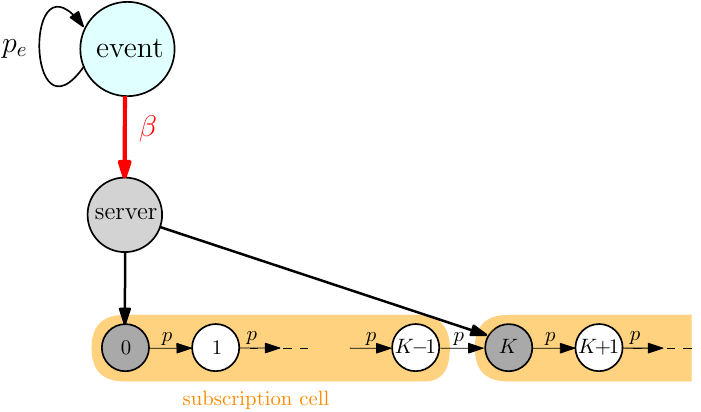}}\vspace*{-0.1cm}
\caption{Directed line network where the users are subscribing to the server's information in every $K$ users.}
\label{fig:directed_line_network}
\vspace*{-0.4cm}
\end{figure}

\subsection{Game Formulation}
We next describe the non-cooperative game between the server and the users. The server wants to maximize the fraction of population that subscribes, denoted by $F_S = \frac{\sum_{i=1}^{|\mathcal{N}|}a_i}{|\mathcal{N}|}$ where $|\mathcal{N}|$ is the cardinality of the user set $\mathcal{N}$, to improve earnings, however, it also wishes to keep $\beta$ low to minimize its cost of tracking the event process. Thus, the server's utility function for given $\beta$ and the action vector $\bm{a}=[a_1,a_2,\cdots, a_{|\mathcal{N}|}]^T$ is given by the expression
\begin{align} \label{eqn:servers_cost}
    L_S(\beta, \bm{a}) = F_S- c(\beta),
\end{align}
where $c(\beta)$ is the sampling cost with rate $\beta$. Here, we consider a general sampling cost $c(\beta)$ that is an increasing function of $\beta$. On the other hand, users take their actions based on their ability to satisfy the AC constraint of (\ref{eqn:ACconstraint}) as discussed earlier. In this game, the server $R$ has the information structure $I_R  = \{\overrightarrow{G}, p_e,p\}$, and the user $i\in \mathcal{N}$ has $I_i  = \{\overrightarrow{G}, p_e,p, \beta, \bm{a}_{-i} \}$,  respectively. Here, we use $\bm{a}_{-i}$ to denote the action of the users other than user $i$. Based on the information structure, server $R$ commits to a strategy $\eta_R :I_R\rightarrow [0,1]$. Similarly, let user $i$'s strategy be $ \gamma_i:I_i\rightarrow\{0,1\}$. More specifically, in this work, we consider user $i$'s action $a_i =\gamma_i(I_i) = \mathbf{1}\{ x_{i,ns}> L x_S \}$ where $x_{i,ns}$ is the average age when user $i$ does not subscribe to the server, for given $\beta$ and $\gamma_{-i}(\beta)$.

Finally, the utility function of server $R$ is given by
\begin{align}
    J_S(\eta_R;\gamma_1, \gamma_2,\ldots, \gamma_{|\mathcal{N}|} ) = L_S(\beta, \bm{a}),
\end{align}
where the dependency of strategies to actions are given by $a_i =\gamma_i(I_i)$ and $\beta =\eta_R(I_S)$. In this game, we consider a hierarchical commitment where first, the server $R$ announces and commits to a strategy $\eta_R$ and then users take action to satisfy the AC condition in (\ref{eqn:ACconstraint}). This structure leads to a Stackelberg game between the server $R$ and the users. We say that the set of strategies $(\eta_R^*,\{\gamma_i^*(\eta_R^*,\gamma_{-i}^*)\}_{i=1}^{|\mathcal{N}|} )$ is in equilibrium if
\begin{align}\label{eqn:eq_pol}
    J_S(\eta_R^*,\!\{\gamma_i^*(\eta_R^*,\!\gamma_{-i}^*(\eta_R^*))\}_{i=1}^{|\mathcal{N}|} \!) &\!\geq\! J_S (\eta_R,\!\{\gamma_i^*(\eta_R,\!\gamma_{-i}^*(\eta_R))\}_{i=1}^{|\mathcal{N}|} )\nonumber\\
    \gamma_i^*(\eta_R^*,\gamma_{-i}^*(\eta_R)) &\!= \!\mathbf{1}\{ x_{i,ns} \geq L x_S \}. 
\end{align}
where user $i$'s subscription strategy will be based on the term $x_{i,ns}$ affected by the server's strategy and other users' subscription strategies. Thus, at the equilibrium, given the understanding of how the users' subscription strategies depend on $\beta$, the server will commit to $\beta^* =\eta_R^*(I_R)$ that will maximize its utility and users adopting $\gamma_i^*$ will not deviate from their action given the server's and other users' subscription strategies. In this work, we will see that depending on the network structure, there might be multiple equilibrium solutions with different subscription structures. In these scenarios, we assume that the users take the server-preferred subscription strategy.\footnote{Another approach could be to consider the worst subscription strategy for the server, and the goal could be to maximize its minimum possible utility.} 

In the next section, we consider directed networks and find the equilibrium subscription strategies.

\section{Directed Networks}\label{sec:directed_networks}
We begin by analyzing equilibrium strategies for directed networks, where information can only flow in one direction between any two nodes $i$ and $j$, that is, $e_{ij}e_{ji}=0$. This analysis not only aids in building a strong intuition, but also provides tools useful for analyzing more complex networks.

\subsection{Directed Line Networks}\label{subsec:directedlinenetwork}
Consider the directed line network shown in Fig.~\ref{fig:directed_line_network}. User $0$ will have to subscribe because not subscribing would lead to an infinite expected age. Let us assume that node $K$ decides to subscribe, where $K>1$, such that nodes $\{1,\ldots,K-1\}$ do not subscribe. Given a specific value of $\beta$, we aim to determine the value of $K$ for which all the users are AC-stable, i.e, any subscribing or non-subscribing user would not wish to change its subscription decision contingent upon the decisions of the other users in the population being fixed. For any node $k\in\{1,\ldots,K-1\}$, similar to (\ref{eqn:instantage_server}) and (\ref{eqn:instantage_subscriber}), we can characterize the instantaneous version age at node $k$ from all combinations of relevant transitions $U_{k-1,k}(t)$ and $U_E(t)$ as follows
\begin{align} \label{eqn:instantage_node_k_directedline}
    X_k(t+1)= &  X_{k-1}(t) \times (1-U_E(t)) U_{k-1,k}(t)\nonumber\\
    & +  X_k(t)\times (1-U_E(t))(1- U_{k-1,k}(t))\nonumber\\
    & +  (X_{k-1}(t)+1)\times  U_E(t)U_{k-1,k}(t)\nonumber\\
    & + (X_k(t)+1) \times U_E(t)(1- U_{k-1,k}(t)),
\end{align}
which, similar to (\ref{eqn:longtermexp_age_server}) and (\ref{eqn:longtermexp_age_subscriber}), yields
\begin{align} \label{eqn:longtermexp_age_node_k_directedline}
    x_k=& x_{k-1}(1-p_e)p+ x_k(1-p_e)(1-p)+  (x_{k-1}+1)p_ep \nonumber\\
    &+ (x_k+1)p_e(1-p).
\end{align}

\begin{figure}[t]
\centerline{\includegraphics[width=0.60\linewidth]{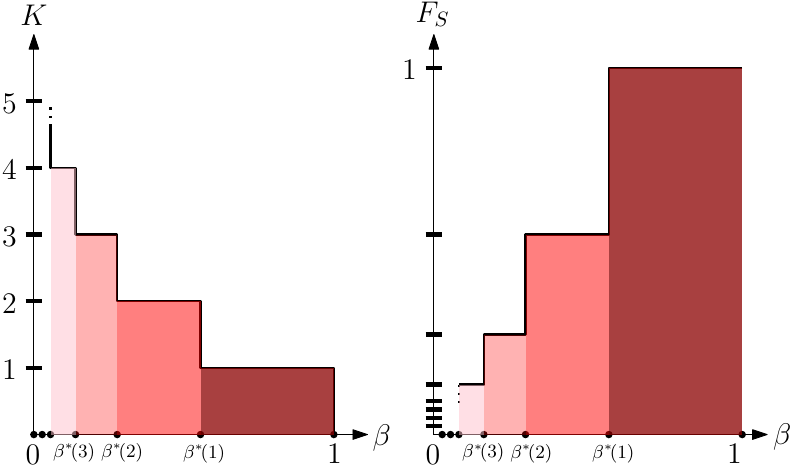}}\vspace*{-0.2cm}
\caption{$F_S$ and $K$ values at different $\beta$.}
\label{fig:FS_K_variation_beta}
\vspace*{-0.4cm}
\end{figure}

From (\ref{eqn:longtermexp_age_node_k_directedline}), we obtain the following recursive relation
\begin{align}\label{eqn:recursive_directedline}
    x_k=x_{k-1}+p_e/p.
\end{align}
Since user $0$ is a subscriber node, from (\ref{eqn:longtermexp_age_subscriber}), we have
\begin{align} \label{eqn:expectedage_S_directedline}
    x_0=x_S=p_e\left(\beta^{-1}+1\right),
\end{align}
and therefore, expected age at node $k$ can be computed by using (\ref{eqn:recursive_directedline}) and (\ref{eqn:expectedage_S_directedline}) as
\begin{align}\label{eqn:xk_expectedage_directedline}
    x_k=x_0+kp_e/p= p_e\left(\beta^{-1}+1\right) + kp_e/p,
\end{align}
such that $x_k$ is inversely proportional to gossiping frequency $p$ and server effort $\beta$. Note that the expected age increases with distance from the previous subscriber node which is node $0$, i.e., $x_1<\ldots< x_{K-1}$, and thus variable $K$ (the index of the next subscriber node after node $0$) must satisfy
\begin{align} \label{eqn:Kcond_directedline}
    x_{K-1} < Lx_S \leq \Tilde{x}_{K},
\end{align}
where we use $\Tilde{x}_{i}$ to denote the expected age at node $i$ if it chooses an alternate subscription decision. The subscriber node $K$ would be AC-stable if the alternative decision to unsubscribe would lead to expected age $\Tilde{x}_{K}=x_0+Kp_e/p$, which would cause it to no longer satisfy the AC constraint of (\ref{eqn:ACconstraint}). For a given $\beta$, this critical value of $K$ from (\ref{eqn:Kcond_directedline}) is
\begin{align}
    &x_S+(K-1)p_e/p < Lx_S \leq x_S+Kp_e/p,
\end{align}
which leads to,
\begin{align}
    K= \left\lceil p(L-1)\left(\beta^{-1}+1\right) \right\rceil, \label{eqn:K_withgivenbeta_directedline}
\end{align}
where $\lceil.\rceil$ denotes the ceil function; see Fig.~\ref{fig:FS_K_variation_beta}. Further, the non-subscriber nodes would also be AC-stable as long as they continue to satisfy the AC constraint, which corresponds to the left-side inequality in (\ref{eqn:Kcond_directedline}). The fraction of the population that subscribes is thus given by
\begin{align}\label{eqn:FS_linedirected}
    F_{S,line}(\beta)= K^{-1} = \left(\left\lceil p(L-1)\left(\beta^{-1}+1\right) \right\rceil\right)^{-1},
\end{align}
which, the server would desire to maximize while maintaining a low $\beta$, where we show the dependency of $F_S$ on $\beta$ by writing $F_S(\beta)$. The minimum $\beta=\beta^*(K)$ required to maintain the subscription frequency $K$ can be given, using (\ref{eqn:longtermexp_age_subscriber}), (\ref{eqn:xk_expectedage_directedline}), and
(\ref{eqn:Kcond_directedline}) by the following relation
\begin{align}
    Lx_S= \Tilde{x}_K= x_S + Kp_e/p,
\end{align}
which yields
\begin{align}\label{eqn:line_critical_beta}
    \beta^*(K)= \left(\frac{K}{p(L-1)}-1\right)^{-1}.
\end{align}
As the server effort $\beta^*$ increases, subscribers appear on the network more frequently, since $K$ decreases, as shown in Fig.~\ref{fig:FS_K_variation_beta}. Next, we use the server's minimum $\beta^*(K)$ to find $\beta^*_{line}$ that maximizes the server's utility $L_S(\beta, \bm{a})$ given in (\ref{eqn:servers_cost}). Thus, the server's optimization problem becomes, 
\begin{align}\label{eqn:server_s_opt_}
    \beta^*_{line} = \argmax_{\beta^*(K),K =1,2,\cdots}  F_{S,line}(\beta^*(K))- c(\beta^*(K)). 
\end{align}
In other words, the server's optimum $\beta^*_{line}$ selection can be found easily by evaluating server's cost at the critical $\beta^*(K)$ values found in (\ref{eqn:line_critical_beta}).  

\begin{figure}[t]
\centerline{\includegraphics[width=0.7\linewidth]{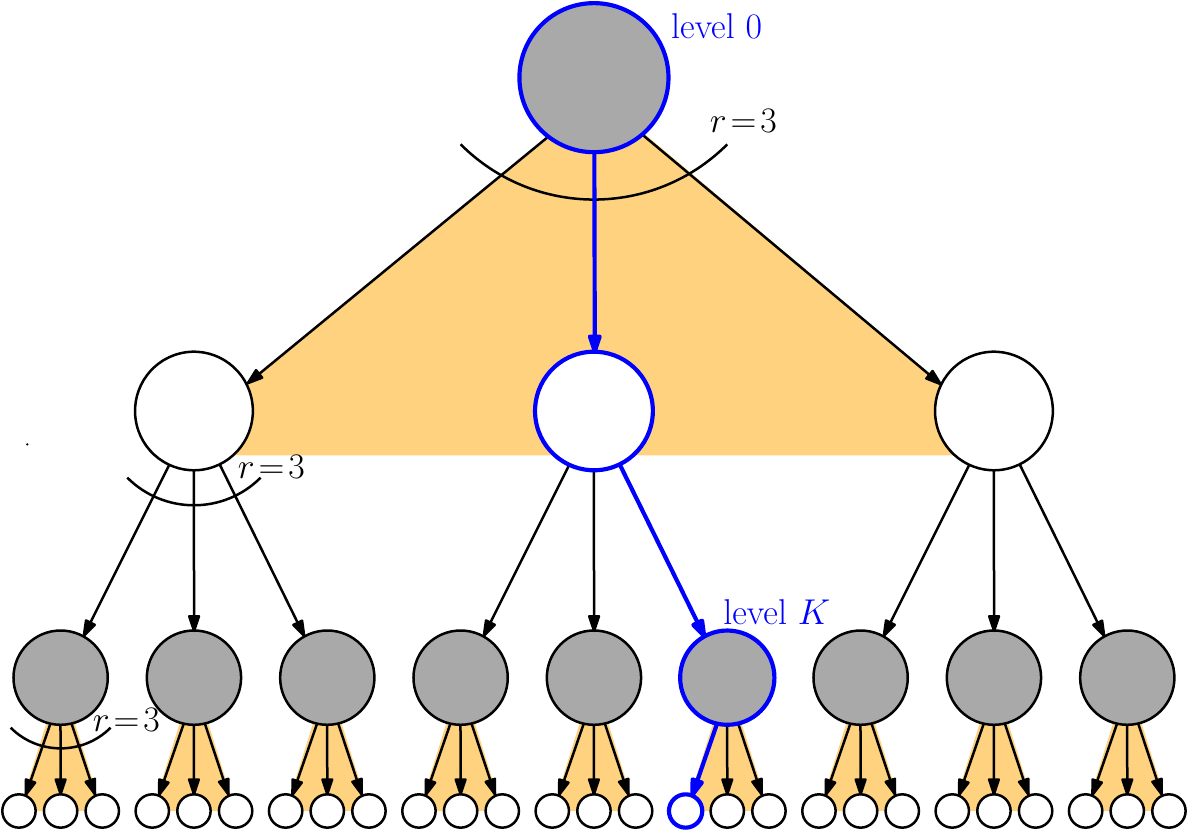}}
\vspace*{-0.2cm}
\caption{Directed tree network with $K=2$ levels.}
\label{fig:tree_network}
\vspace*{-0.4cm}
\end{figure}

\begin{figure*}[t]
\centerline{\includegraphics[width=0.6\linewidth]{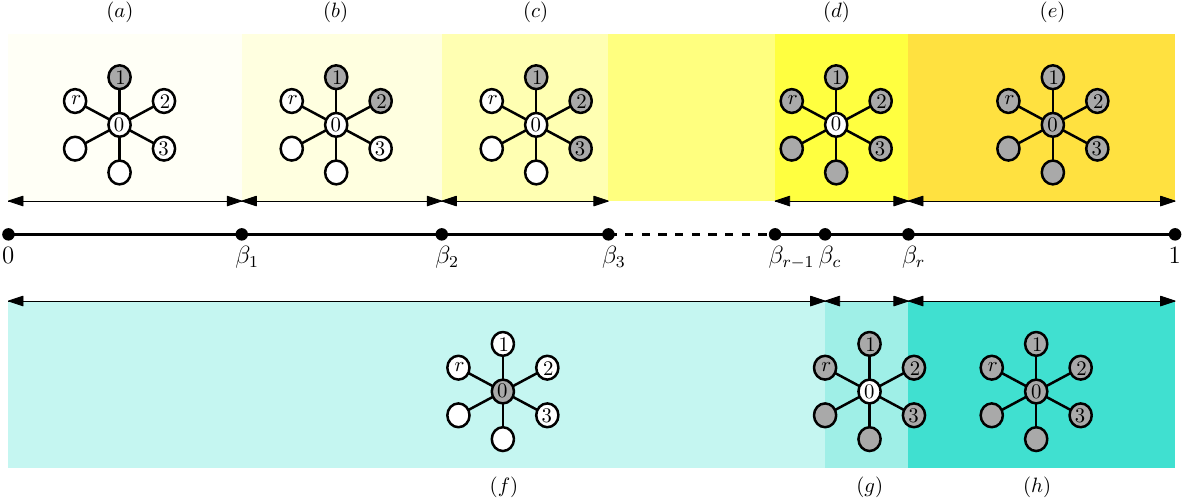}}\vspace*{-0.4cm}
\caption{Equilibrium subscription strategies on star network as function of 
$\beta$.}
\label{fig:star_network}
\vspace*{-0.4cm}
\end{figure*}

\subsection{Directed Tree Networks}
Next, we focus on the network topology of Fig.~\ref{fig:tree_network}, where users are connected according to an $r$-regular tree, such that each user can transmit updates to $r$ new users, $r>1$. In the tree network, therefore, there are $r^k$ nodes at level $k$; see Fig.~\ref{fig:tree_network}. The root node here must be a subscriber because not subscribing would result in an infinite expected age. A tree network provides a more realistic representation of how information spreads in a society. Here, each directed path in the tree is simply a directed line network of Subsection~\ref{subsec:directedlinenetwork}; therefore, the nodes in levels $\{1,\ldots,K-1\}$ will not subscribe, and all nodes at level $K$ will subscribe, where $K$ is given by (\ref{eqn:K_withgivenbeta_directedline}), i.e., the whole level subscribes at intervals of $K$; see subscription cells of Fig.~\ref{fig:tree_network}. The number of users in each subscription cell of Fig.~\ref{fig:tree_network} is $r^0+r^1+\ldots+r^{K-1}=\frac{r^K-1}{r-1}$, and therefore, the fraction of the subscribers is given by
\begin{align}\label{eqn:FS_treedirected}
    F_{S,tree}(\beta)=\frac{r-1}{r^K-1} = \frac{r-1}{r^{\left\lceil p(L-1)\left(\beta^{-1}+1\right) \right\rceil}-1}.
\end{align}

From (\ref{eqn:K_withgivenbeta_directedline}) and (\ref{eqn:FS_treedirected}), we note the inverse relationship between $F_{S,tree}$ and $\beta$. Comparing (\ref{eqn:FS_linedirected}) and (\ref{eqn:FS_treedirected}), we see that with the same server effort $\beta$, we get fewer subscribers in a tree network due to the exponential decay of (\ref{eqn:FS_treedirected}), compared to when users are connected in a line network. The tree network has one user directly updating $r$ other users, and the server needs to increase $\beta$ to stay relevant for such well-connected communities. After characterizing the AC-stable user's subscription strategies for a given $\beta$, next we find the optimal $\beta^*_{tree}$ that will maximize the server's utility. Thus, the server's optimization problem in tree networks is given by 
\begin{align}\label{eqn:server_s_opt_tree}
    \beta^*_{tree} = \argmax_{\beta^*(K),K =1,2,\cdots}  F_{S,tree}(\beta^*(K))- c(\beta^*(K)). 
\end{align}
The server's optimal $\beta^*_{tree}$ can be found by evaluating the server's cost function at $\beta^*(K)$ values for $K\geq 1$.  

\section{Star Networks}
We next evaluate the star network topology of Fig.~\ref{fig:star_network}, where a central user (node $0$) is connected to $r$ peripheral users. This is an undirected network, that is, information flow is bi-directional on all edges of  $\overrightarrow{G}$, however, for any given strategies $\gamma_1, \gamma_2,\cdots, \gamma_{|\mathcal{N}|}$, this becomes a directed network. 
Let us denote the age of a peripheral non-subscribing user by $x_{NP}$. Since a peripheral non-subscriber gets updates from the central user, using (\ref{eqn:recursive_directedline}) we get
\begin{align}\label{eqn:nonperipheralsubscriberage}
    x_{NP}=x_0+p_e/p.
\end{align}

We first consider strategies in which the central user is a subscriber. When the central user subscribes, $x_0=x_S$. Suppose that there are $k$ peripheral subscribers along with the central user in the network. The remaining $r-k$ peripheral non-subscribers will be AC-stable, if each of them satisfies the AC constraint. Using (\ref{eqn:longtermexp_age_subscriber}), (\ref{eqn:ACconstraint}), and (\ref{eqn:nonperipheralsubscriberage}), this gives
\begin{align}\label{eqn:periph_NS_centr_subs}
     x_S + p_e/p< Lx_S,
\end{align}
which in turn yields
\begin{align}\label{eqn:beta_c_centralnode}
    \beta < \left(\frac{1}{p(L-1)}-1\right)^{-1}=\beta_c.
\end{align}
However, if $\beta<\beta_c$, any peripheral subscriber will not be AC-stable, since a peripheral subscriber can change its decision and unsubscribe, which will still satisfy the AC constraint from (\ref{eqn:periph_NS_centr_subs}). This results in the user strategies of Fig.~\ref{fig:star_network}(f), where given $\beta<\beta_c$, only the central user subscribes and all peripheral users do not subscribe. In this case, the central user is AC-stable since not subscribing would result in an infinite age. 

If $\beta\geq\beta_c$, and we still consider that the central node is a subscriber, then no peripheral non-subscribers can be AC-stable as the AC constraint of (\ref{eqn:periph_NS_centr_subs}) will not be satisfied. Hence, all peripheral nodes must subscribe. However, to make sure that the central user is also AC-stable, we need to verify that the alternate age $\Tilde{x}_0$, when the central user chooses the alternate decision to not subscribe, violates the AC constraint. 
Approaching similar to (\ref{eqn:instantage_server}) and (\ref{eqn:instantage_subscriber}), we get
\begin{align}
    \Tilde{x}_0=\frac{p_e}{1-(1-p)^r}+x_S.
\end{align}
For $\Tilde{x}_0$ to violate the AC constraint ($\Tilde{x}_0 \geq Lx_S$), we require
\begin{align} \label{eqn:beta_r_centralnode}
    \beta \geq \left(\frac{1}{(1-(1-p)^r)(L-1)}-1\right)^{-1} = \beta_r.
\end{align}
Since $p =1-(1-p)  \leq 1-(1-p)^r$, we have from (\ref{eqn:beta_c_centralnode}) and (\ref{eqn:beta_r_centralnode}) that $\beta_c \leq \beta_r$. Thus, for $\beta \geq \beta_r$, we get the user subscription strategies of Fig.~\ref{fig:star_network}(h), where all users subscribe. When $\beta \!\in\! [\beta_c,\beta_r)$, we will see in the following discussion that the subscription strategies of Fig.~\ref{fig:star_network}(g) follow, where all the $r$ peripheral users subscribe, but the central user does not.

We now look at user strategies where the central user is not a subscriber. Suppose that there are $k$ peripheral subscribers with $k\!\geq\! 1$ since $k\!=\!0$ will lead to infinite age for all users. We find the values of $\beta$ which guarantee that the non-subscribing and subscribing users will be AC-stable, which yield, respectively, an upper bound and a lower bound on $\beta$ as follows
\begin{align}
  \beta_{k-1}  \leq \beta < \beta_k, \quad 1\leq k \leq r,
\end{align}
where
\begin{align}\label{eqn:beta_k-1_star}
    \beta_{k} =  \begin{cases} 
      \frac{1}{\left(\frac{1}{1-(1-p)^{k}}+\frac{1}{p}\right) \frac{1}{L-1}-1}, & 1 \leq k \leq r-1, \\
      \frac{1}{\frac{1}{(1-(1-p)^r)(L-1)}-1}, & k = r,
   \end{cases}
\end{align}
such that $\beta_k$ is an increasing function of $k$. Note that $\beta_0=0$ as $k=0$ peripheral subscriber could get infinite age upon unsubscribing. Therefore, when $\beta\in[\beta_{k-1},\beta_k)$, $k=1,\ldots,r$, $k$ peripheral users will subscribe and all other nodes will not subscribe, as shown in Figs.~\ref{fig:star_network}(a)-(d). As $\frac{1}{p}\! <\! \frac{1}{1\!-\!(1\!-\!p)^{k\!-\!1}}\!+\!\frac{1}{p}$, comparing (\ref{eqn:beta_c_centralnode}) and (\ref{eqn:beta_k-1_star}), we get $\beta_{r-1}< \beta_c$. Therefore, the critical values of $\beta$ can be ordered as\footnote{Given the overlap of Fig.~\ref{fig:star_network}(f) with strategies of Fig.~\ref{fig:star_network}(a)-(d) for $\beta < \beta_c$, if a news agency plans to advertise promotional offers to increase the number of its subscribers despite a low $\beta$, it should target peripheral users with low connectivity instead of the central user, since subscription by the central user disincentivizes a large number of users from subscribing.
}
\begin{align}
0<\beta_1<\ldots<\beta_{r-1}< \beta_c<\beta_r<1.
\end{align}
Since we assume that the users take the server-preferred subscription strategy, the subscription ratio is equal to
\begin{align} 
F_{S,star}(\beta) = \begin{cases} 
\frac{k}{r+1}, & \beta \in [\beta_{k-1},\beta_k),~ k=1,\ldots,r,\\
1, & \beta \geq \beta_r.
\end{cases}
\end{align}
The minimum $\beta=\beta^*(k)$ required to maintain the $k$ subscribers can be given by $\beta^*(k)=\beta_{k-1}$. 
After characterizing the AC-stable user's subscription strategies for a given $\beta$, next, we find the optimal $\beta^*_{star}$ maximizing the server's utility. Thus, the server's optimization problem in star networks is given by
\begin{align}\label{eqn:server_s_opt_tree}
    \beta^*_{star} = \argmax_{\beta_{k-1},k =1,2,\cdots,r+1}  F_{S,star}(\beta_{k-1})- c(\beta_{k-1}). 
\end{align}
The optimal $\beta^*_{star}$ can be found by evaluating server's cost at $\beta_{k-1}$ values for different numbers of subscribers $k$.  

\begin{figure}[t]
\centerline{\includegraphics[width=0.65\linewidth]{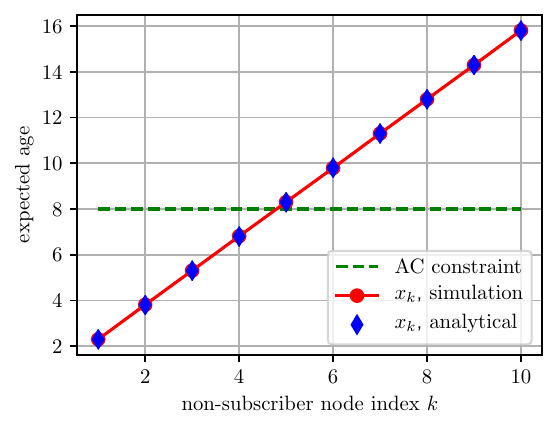}}
\vspace*{-0.1cm}
\caption{Age evolution in one-way line network when $L=10$, $p=0.2$, $\beta=0.6$, and $p_e=0.3$. For all users to be AC-stable, users in $5\mathbb{N}$ will have to subscribe.}
\label{fig:graph_xk_vs_k_onewayLine_numerical_VS_analytical}
\vspace*{-0.4cm}
\end{figure}

\section{Simulations}
We first simulate the line network in Fig.~\ref{fig:graph_xk_vs_k_onewayLine_numerical_VS_analytical} with parameters $L=10$, $p=0.2$, $\beta=0.6$, and $p_e=0.3$ for a total time of $t=10^4$, which we use as a proxy for $t \to \infty$, and take the average over $2\times10^5$ iterations to approximate $\lim_{t \to \infty} \mathbb{E}[X_i(t)]$ by the law of large numbers. The real-time simulation points (red) coincide with the analytically derived values (blue) from (\ref{eqn:xk_expectedage_directedline}), verifying the theoretical analysis. The graph shows how expected age increases in a directed path linearly with the number of hops from the nearest subscriber; thus, we have $K=5$ for all users to be AC-stable, as in (\ref{eqn:K_withgivenbeta_directedline}).

\begin{figure}[t]
\centerline{\includegraphics[width=0.65\linewidth]{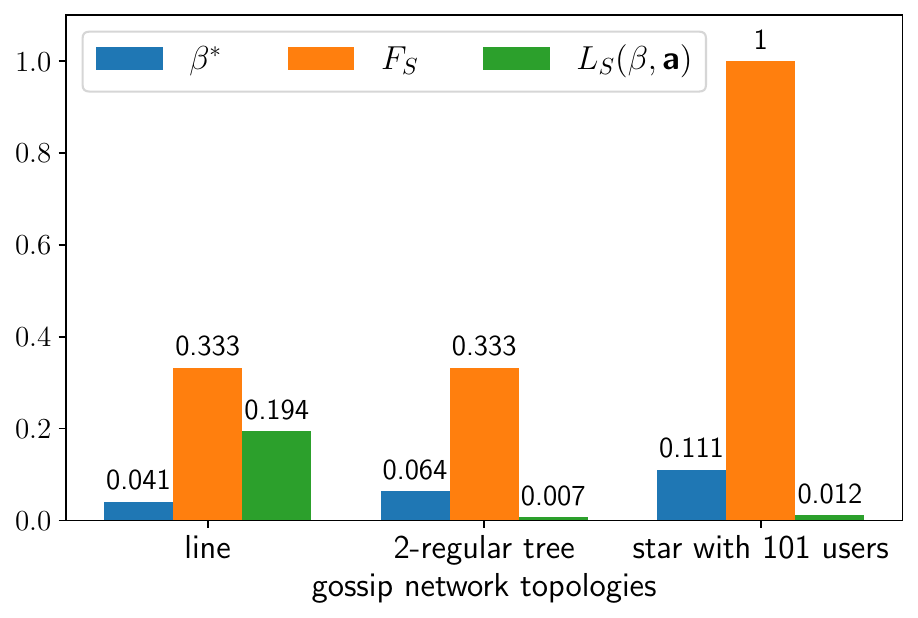}}
\vspace*{-0.2cm}
\caption{The comparison of the server's optimal $\beta^*$, subscription rate $F_S$, and the server's optimal cost function for the line, tree, and star networks when $L=10$, $p=0.2$, $\beta=0.6$, and $p_e=0.3$.}
\label{fig:graph_graph_beta_FS_LS_directed}
\vspace*{-0.4cm}
\end{figure}

Next, we analyze the equilibrium strategies for the server and the users in a line network, $2$-regular tree network, and a star network with $100$ peripheral users, with parameters $L=10$, $p=0.2$, $\beta=0.6$, and $p_e=0.3$, for the server utility function $F_S(\beta)- 80 \times\beta^2$. This utility function penalizes high values for $\beta$, but is indifferent to smaller values of $\beta$ due to the $80 \times\beta^2$ term. Fig.~\ref{fig:graph_graph_beta_FS_LS_directed} shows $\beta^*$ and the corresponding $F_S$ and server utility function $L_S(\beta, \bm{a})$ achieved in all three networks. We note that a higher $\beta^*$ is required in tree network to achieve the same $F_S$, and the corresponding server utility is lower since nodes are more well-connected in a tree. On the other hand, in star network, we choose the highest $\beta^*=0.111$ that causes all users to subscribe; however, the high $\beta^*$ negatively affects the server utility, which is the lowest for star network.

\section{Conclusion}
We investigated a Stackelberg game between a server and users, where the server wishes to maximize its profit by increasing the number of subscribers and reducing costs associated with the frequent sampling of the event, while the users make their subscription decisions based on their timeliness requirements. We analyzed directed networks and showed how subscribers repeat periodically along directed paths in line and tree networks. Our theoretical results revealed an inverse relationship between the expected age and both gossiping and server sampling rates. The analyses in tree and star networks showed how higher connectivity and the presence of central agents can discourage other nodes from subscribing, and well-connected communities tend to have fewer subscribers.

\bibliographystyle{unsrt}
\bibliography{ref_priyanka}

\end{document}